\documentstyle[12pt,aas2pp4]{article}
\def\insfig#1{#1}
\def\endinsfig{\end{document}}

\font\smallrm=cmr8

\def\FWHM{{\smallrm FWHM}}

\def\kms{\hbox{$\,$km$\,$s$^{-1}$}}
\def\kmsMpc{\hbox{$\,$km$\,$s$^{-1}\,$Mpc$^{-1}$}}
\def\kpc{\hbox{$\,$kpc}}
\def\Mpc{\hbox{$\,$Mpc}}

\lefthead{Tonry\&Franx}
\righthead{Dispersion of 0957}

\begin{document}

\title{Velocity Dispersion of the Gravitational Lens 0957+561\altaffilmark{1}}

\author{John L. Tonry}
\affil{Institute for Astronomy, University of Hawaii, Honolulu, HI 96822}
\affil{Electronic mail: jt@avidya.ifa.hawaii.edu}
\authoremail{jt@avidya.ifa.hawaii.edu}

\author{Marijn Franx}
\affil{Kapteyn Astronomical Institute, P.O. Box 800, NL-9700 AV,
Groningen, The Netherlands}
\affil{Leiden Observatory, P.O. Box 9513, NL-2300 RA,
Leiden, The Netherlands}
\affil{Electronic mail: franx@strw.leidenuniv.nl}
\authoremail{franx@strw.leidenuniv.nl}

\altaffiltext{1}{Based on observations at the W. M. Keck Observatory,
which is operated jointly by the California Institute of Technology
and the University of California}

\begin{abstract}
0957+561 is the first gravitational lens system to be discovered and
first for which a time delay was measured.  Because the system is
unusually rich in observables such as image positions, fluxes, VLBI
structure, and polarization it has been modelled quite extensively.
However, since it resides in a cluster providing substantial
convergence, the velocity dispersion of the
lensing galaxy is an important component to models.  We have measured
this dispersion to be $288\pm9$\kms\ (1-sigma), which implies a
Hubble constant of $72\pm7$\kmsMpc\ (1-sigma) according to the
preferred FGS model of Grogin and Narayan, and
$70\pm7$\kmsMpc\ (1-sigma) using the SPLS model.
The unknown velocity dispersion anisotropy of the central galaxy
produces an additional uncertainty of perhaps 15\% or more.
We see no variation in dispersion with position
to a radius of $\pm3$\arcsec.  
In addition to a galaxy from the
background cluster at $z=0.5$, our slit serendiptously picked up 
a galaxy at $z = 0.448$, and two galaxies which may be part of a
background cluster at a redshift of $z=0.91$.

\noindent \it{Subject headings:} cosmology --- distance scale ---
gravitational lensing --- quasars: individual (0957+561)

\end{abstract}

\section{Introduction}

Gravitational lenses bend light from a background source and can cause
multiple images of the source to appear.  Light has a different time
of flight along each path, and if the source should vary there will be
a time delay between when each of the images changes brightness.  Such
a time delay can transform dimensionless redshifts and angles into
physical distances, hence provide a Hubble constant (Refsdal 1964).
In principle this is an extremely powerful and important method for
measuring the cosmological parameters embodied in distances and
redshifts, party because of the potential accuracy at high redshift
but mainly because they suffer a completely different set of biases
from more traditional measures of $H_0$.

A time delay for the lensed system 0957+561 has been known for a
decade (Schild and Cholfin 1986), although Press, Rybicki, and Hewitt
(1992) disputed their value and only recent measurements by Kundi\'c
et al. (1997) have settled the issue beyond doubt.  See Haarsma et
al. (1997) for a review of the time delay measurement.  Models of the
system have also undergone improvement; Grogin \& Narayan (1996) [GN]
have carried out a particularly extensive analysis which has been
widely used for estimating $H_0$.  Romanowsky \& Kochanek (1998) have
recently examined the question of how velocity anisotropy in the
lensing galaxy affects the use of a central velocity
dispersion to estimate the lensing potential.

The models of GN can yield a Hubble constant either by using an
estimate of the local convergence $\kappa$ due to the cluster potential 
or by measuring the depth
of the lensing galaxy potential from its velocity dispersion
$\sigma$.  The first measurement of the velocity dispersion was by
Rhee (1990) who found $300\pm50$\kms.  Recently, observations with
LRIS on the Keck telescope by Falco et al. (1997, FSMD) have found a
dispersion which drops rapidly with radius, from $316\pm14$ at
0.2\arcsec\ from the center of the lensing galaxy G1, dropping to
$266\pm12$ at 0.4\arcsec.  Using the latter value they derive a Hubble
constant of $H_0 = 62\pm7$\kmsMpc, and they speculated that a 
massive central object might be causing an anomalously large
central dispersion.

The very rapid drop in velocity dispersion with distance is unusual
behavior for such a massive galaxy, however, which inspired an attempt
to reobserve the lensing galaxy in 0957.  One concern about the
observations of FSMD centers on their choice to run the slit through
the G1 galaxy and the QSO B component simultaneously.  While this
offers both the galaxy and QSO light in one exposure it suffers from
much more scattered light and could potentially corrupt the galaxy
spectrum in unforseen ways.  Indeed, FWMD suggest redoing their
observation with the spectrograph slit rotated by 90 degrees.

We chose to do just that: rotate the slit perpendicular to
the line connecting G1 and the QSO B component, excluding as much QSO
light as possible.  In addition, we attempted to match the QSO
contribution leaking into the slit as accurately as possible by taking
alternate exposures with the slit rotated by $180^\circ$ and offset to
the other side of the QSO B component, thus obtaining a relatively
pure QSO spectrum which illuminates the slit in the same way as during
the G1 exposure.

\section{Observations and Reductions}

0957+561 was observed on March 30 and March 31, 1997
using the Low Resolution Imaging Spectrograph (LRIS) (Oke et al. 1995)
at the Keck II telescope on Mauna Kea, along with calibration
exposures, an observation of the cluster MS1358+62 to act as a
velocity dispersion calibrator, and HD132737 and AGK2+14873 as radial
velocity templates.  The observations are summarized in Table 1.  The
sky was clear and the seeing was about 0.8\arcsec\ throughout both
nights.  Long slits of 1.0\arcsec\ and 0.7\arcsec\ were used along
with the 600~l/mm grating blazed at 5000\AA.  The
600~l/mm grating was rotated either to
cover 4800--7680\AA\ for the QSO and cluster observations or 3800--6080\AA\ for
the template stars.  The spectral resolution was 
4.65\AA\ \FWHM\ for the 600~l/mm grating with a
1.0\arcsec\ slit, 3.59\AA\ \FWHM\ for the 600~l/mm grating with a
0.7\arcsec\ slit,  and the scale along the slit was
0.211\arcsec/pixel.  The template stars were guided smoothly across
the slit in many locations in order to have uniform illumination
across the slit, to build up signal to noise, and to map out loci of
constant slit position across the detector.  

Figure 1, derived from a very deep image by Fischer et al. (1997), 
illustrates where the three slit positions lay.  PA108 runs through
the lensing galaxy, G1, two neighboring galaxies, t2 and t3, and also
catches two other galaxies at the end of the slit, t1 and t4.  Since
we needed as clean a spectrum of G1 as possible we rotated the slit by
$180^\circ$ to PA288 and offset to a point symmetrically located on
the other side of QSO image B.  This also yielded a spectrum of
another galaxy, t5 as well as t1 again.  Finally, we offset to center
the slit on QSO image A to acquire a pure QSO image, and we also
picked up galaxies t6, t7, and t8.

\insfig{
\begin{figure}[t]
\epsscale{1.0}
\plotone{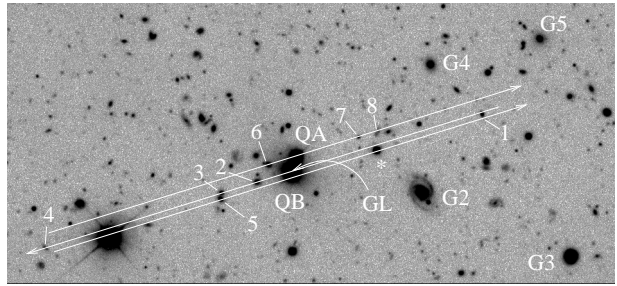}
\caption[f1.eps]{
Illustration of the slit positions and identifications of the galaxies
observed here.  North is up and East is left; G2--G5 are the
designations of Young et al. (1980); and t6 is galaxy 20 and t2 is
galaxy 21 from Angonin-Willaime et al. (1994).
\label{fig1}}
\end{figure}
}

The spectra were reduced using software described in detail by Tonry
(1984).  The basic steps are to flatten the images, remove cosmic
rays, derive a wavelength solution as a function of both row and
column using sky lines (wavelengths tabulated by Osterbrock et
al. 1996), derive a slit position solution as a function of both row
and column using the positions of the template star images in the
slit, rebin the entire image to coordinates of log wavelength and slit
position, add images, and then sky subtract.  A quadratic fit to
patches of sky on either side of the object (including a patch between
for the MS1358 observations) did a very good job of removing the
sky lines from the spectra.

As has been stressed by Franx (1993) and by van Dokkum and Franx (1996),
measuring the velocity
dispersion of a galaxy at a redshift significantly larger than zero
must be done carefully, since the instrumental resolution of a
spectrograph tends to a constant number of angstroms, whereas the
redshifted spectrum has been stretched.  Hence a straightforward
cross-correlation or fitting method  will underestimate the
dispersion of the galaxy.  Kelson et al. (1997) measured dispersions in the
cluster MS1358+62 using high resolution template spectra and very
careful modelling of the spectrograph resolution derived by measuring
sky line widths.  The observations here use the trick that the ratio
between an 0.7\arcsec\ and 1.0\arcsec\ slit is slightly larger than
$(1+z)$ for these lenses, hence a {\it template} measured with the
0.7\arcsec\ slit will have almost exactly the same instrumental
resolution as a {\it galaxy} at a redshift of $z\approx0.3$.  This is
borne out by the ratio of the measured spectral resolutions: 
4.65\AA\ $\div$ 3.59\AA\ = 1.30.

As a test that this observing procedure will give correct dispersions,
we also observed the cluster MS1358+62 for which Kelson et al. (1997)
measure a dispersion of $303\pm6$ for the central galaxy, G1, and
$219\pm4$ for the neighboring bright galaxy, G2.  We discuss our
observations of MS1358 in Tonry (1998), but the salient point is that
the velocity dispersions measured for the MS1358 galaxies using the
stellar templates reveals no significant offset from Kelson's values,
G1: $299\pm22$; G2: $235\pm23$.  We therefore conclude that
use of the stellar templates is likely to give us accurate velocity
dispersions.

\subsection{Companion Galaxies}

Extraction of the spectra of the companion galaxies required no 
correction for QSO light.  In each case the spectrum was extracted, and
cross-correlated with the template spectrum according to Tonry and
Davis (1979) as well as being analyzed by the Fourier quotient method
of Sargent et al. (1977).  For
each spectrum the redshift, error, and velocity dispersion were
calculated.  The redshift calculation included the entire spectrum,
whereas the spectrum blueward of 4000\AA\ (rest frame) was excised for
the dispersion calculation, since the calcium H and K lines,
intrinsically broad and subject to interstellar absorption, 
are always problematic for dispersions.  Dispersions
are not corrected for any aperture effects, hence correspond to an
aperture of approximately 1\arcsec\ square.

Three galaxies, t4, t7, and t8, showed unmistakable emission lines at
7126, 7176, and 7105\AA\ respectively, and t7 also had a strong emission
line at 5396\AA, as well as ones which might be present at 7037\AA\ and
7248\AA, although they are confused with night sky lines.
We believe that the emission in t4 is [OII] 3727\AA\
because of the corroboration from absorption lines at the red end of
the spectrum which correspond well to the UV lines between [OII] and
Ca K.  Galaxies t7 and t8 only received an integration of 500 sec, so
it is more difficult to be certain of their redshift.  The four
emission lines in t7 correspond well to [OII] at 5376\AA\ and
$H_\beta$ and the [OIII] lines in the red, so we are quite confident
about the redshift we assign it.  Galaxy t8 is much more problematic,
with only one emission line present.  We tentatively identify this
line with [OII], giving it a redshift similar to that of galaxy t4,
but a deeper integration will
be necessary before we have much confidence in this redshift.

Table 2 lists the redshifts, errors, velocity dispersions, errors, and
cross-correlation significance ``$r$'' values for each spectrum.  The
spectra are illustrated in Figure 2.
Comparison of the redshifts here with those measured by Garrett et
al. (1992) and Angonin-Willaime et al. (1994) shows complete agreement
within the errors.

\insfig{
\begin{figure}[tbph]
\epsscale{1.0}
\plotone{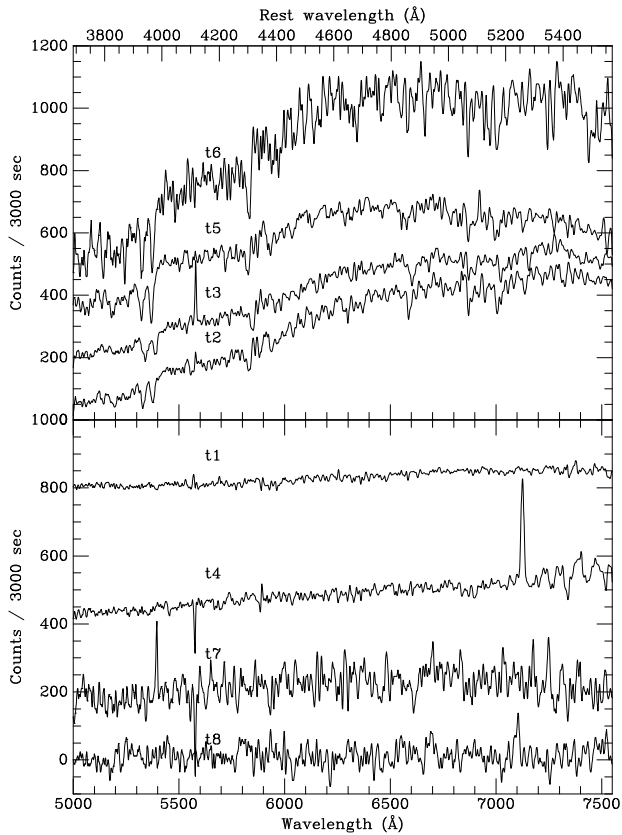}
\caption[f2.eps]{ The spectra are shown for the four galaxies at
redshift 0.355, t6, t5, t3, and t2 (offset by 400, 250, 150, and 0 for
clarity), and the four galaxies at higher redshift, t1, t4, t7, and t8
(offset by 800, 400, 150, and 0).
These spectra have been Gaussian smoothed with a \FWHM\ of
6\AA.  The top axis shows rest wavelength at a redshift of 0.355.
\label{fig2}}
\end{figure}
}

\subsection{Lensing Galaxy}

Anticipating a serious contamination of the spectrum of the lensing
galaxy G1 by QSO component B, we had collected light from the other
side of B, with the slit arranged so that the light gradient across
the slit would match the exposures of G1.  However, the seeing was
good enough that there was surprisingly little contamination (less
than 30 percent, and the QSO light is mostly featureless), and on
the night of Mar 31 we dispensed with the matching exposures of B.  A
simple scaling of the exposure of QSO component A removed the narrow
absorption lines of MgII and gave us the ``pure'' spectrum of G1
illustrated in Figure 3. The spectrum has been averaged over 11
rows, equivalent to 2.3 arcsec. The effective aperture is
therefore 1 arcsec by 2.3 arcsec.

\insfig{
\begin{figure}[t]
\epsscale{1.0}
\plotone{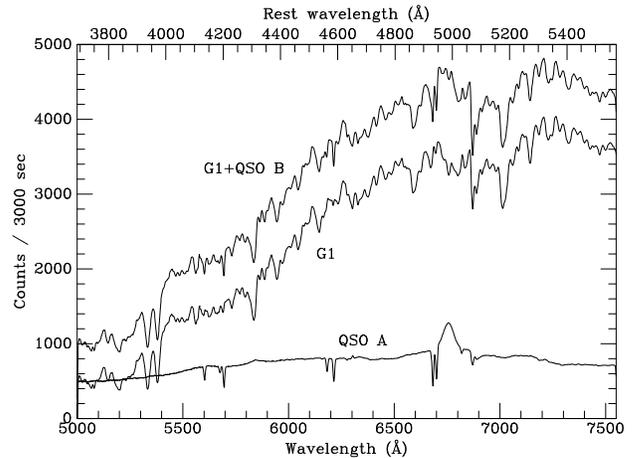}
\caption[f3.eps]{The combined spectrum of the lensing galaxy G1 and
the leakage of light from QSO component B are shown, as well as the
scaled spectrum of QSO component A and the difference which is the
``pure'' spectrum of G1 which is subjected to redshift and dispersion
analysis.  
These spectra have been Gaussian smoothed with a \FWHM\ of
6\AA.  The top axis shows rest wavelength at a redshift of 0.355.
\label{fig3}}
\end{figure}
}

We used three different methods for computing the velocity dispersion
of the lensing galaxy G1.  The first is the standard analysis which 
uses the template star observations.  The second method uses the
MS1358 galaxies as velocity dispersion templates.  The third method
involves fitting the sky lines to determine the instrumental profile
and then creating a synthetic template from higher resolution data.

Using a star as a velocity dispersion template requires a narrower
slit to make the instrumental resultion commensurate for template and
galaxy.  As demonstrated by the comparison between the dispersions
measured for MS1358-G1 and MS1358-G2 with the higher resolution data
of Kelson et al., our trick of using an 0.7\arcsec\ and 1.0\arcsec\
slit is likely to be pretty good at the redshift of 0957, but we
expect that there may be systematic differences in dispersion
according to which wavelengths are compared.  We found no difference
between the four observations of G1, so report only the average
dispersion which we find to be $290\pm20$.  Using the blue half of the
template gave a dispersion of 317 and the red half gave 285, so we
view this result with some caution.

The second method exploits the MS1358 data again, cross-correlating
the lensing galaxy against each of MS1358-G1 and MS1358-G2.  Then,
cross-correlation of MS1358-G1 against MS1358-G2 with varying amounts
of extra broadening allows us to calibrate the width of the
correlation peak of each as a function of velocity dispersion since we
know the dispersions of each of these galaxies.  The advantage here is
that 0957 is at a small redshift relative to MS1358, so the difference
in instrumental resolution is negligible.  The disadvantage is that
this result uses low signal-to-noise templates and depends on the
accuracy of Kelson's dispersions.  Using MS1358-G1 as a template, we
find a dispersion of $285\pm20$ for 0957-G1, and using MS1358-G2 as
the template, we find a dispersion of $283\pm23$.

\insfig{
\begin{figure}[t]
\epsscale{1.0}
\plotone{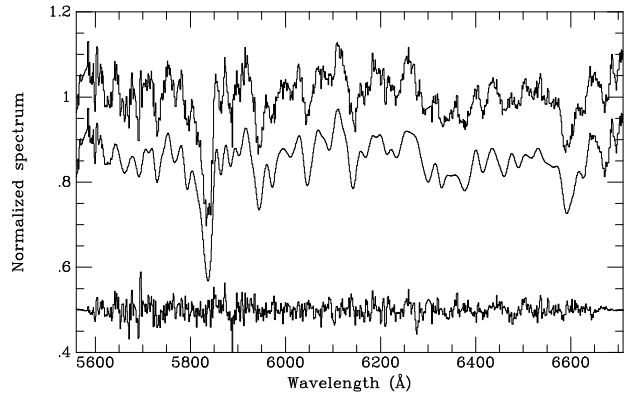}
\caption[f4.eps]{The normalized spectrum, the fit derived from the 
Fourier fit (offset by $-0.15$), and the difference (offset by +0.5)
are plotted as a function of observed wavelength.  The big feature at
5840\AA\ is the G band.  The quality of the fit is clearly very good.
\label{fig4}}
\end{figure}
}

Finally, we use the method introduced by van Dokkum and Franx (1996) 
and measure the width of sky lines, using them to broaden a
synthetic, redshifted, high resolution stellar template which we 
provide to the
Fourier fitting method of Franx et al. (1989).  
Several template stars were fitted to the galaxy spectrum, and only
the best fitting template was used to derive the final velocity
dispersion, and we fitted over the wavelength region of 5558-- 6710\AA\
to avoid the atmospheric B band at 6800\AA.  This reduction also
reproduces Kelson's dispersions of MS1358 very accurately and gives a
velocity dispersion of $289\pm9$ for 0957-G1.
The spectrum and the fit are shown in Figure 4. The quality of the
data and the fit are clearly excellent.
 We regard this as the
most reliable method, but since the other two methods are consistent,
we offer a weighted average of $288\pm9$ as our best estimate for the
velocity dispersion of the gravitational lens.

The dispersion as a function of position along the
slit is illustrated in Figure 5.  
Using the stellar templates, we 
find values of $286\pm30$, $340\pm20$, $294\pm10$, $293\pm10$,
$287\pm10$, $308\pm15$, and $291\pm30$ at positions of $-3.2$\arcsec,
$-1.8$\arcsec, $-0.7$\arcsec, $0.0$\arcsec, $+0.7$\arcsec,
$+1.8$\arcsec, and $+3.2$\arcsec\ along the slit, and using 
the sky line calibration we find values of
$311\pm 28$, $293\pm14$, $286\pm11$, $301\pm14$, and $248\pm24$ at
positions of $-1.7$\arcsec, $-0.6$\arcsec, $0.0$\arcsec,
$+0.6$\arcsec, and $+1.7$\arcsec,
i.e. no significant variation of dispersion with radius.
The discrepancy between the two reduction methods is reasonably
consistent with the formal errors, and the two points at $\pm3$\arcsec\
should not be taken very seriously.

\insfig{
\begin{figure}[t]
\epsscale{1.0}
\plotone{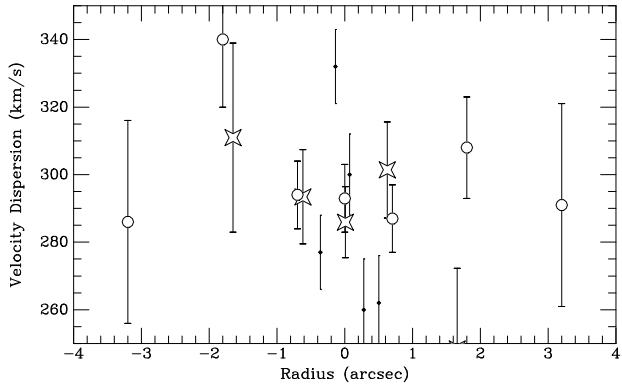}
\caption[f5.eps]{The velocity dispersion is plotted as a function of
position along the slit (open symbols are results from use of stellar
templates; star symbols show results from line fitting).  The data
points from FSMD are also shown as small, filled symbols.
\label{fig5}}
\end{figure}
}

\section{Discussion}

Our redshift for galaxy t1 agrees well with the measurement by Garrett
et al. (1992), confirming the existence of the cluster they found at
$z = 0.5$.  If the redshift of galaxy t8 is confirmed to be 0.907, it
suggests that galaxies t4 and t8 might be part of a background
cluster along the line of sight.  The spectrograph slit was about
160\arcsec\ long and we would probably detect a galaxy within
$\pm1$\arcsec of the slit, so the total area surveyed by the three
slits was about 500 square arcsec.  The angular size distance at
$z=0.91$ is $1355\,h_{75}^{-1}$\Mpc\ (for $\Omega_0 = 0.25$), hence
2 galaxies at that distance projected within 0.02\Mpc$^2$,
is a remarkably high density.
The rest frame velocity dispersion suggested by the two galaxies is
about 800\kms.

The velocity dispersion derived here differs somewhat from that
reported by FSMD.  As shown in Figure 5, they found a dispersion which
dropped very rapidly from 316\kms\ at 0.2\arcsec\ =
$0.86\,h_{75}^{-1}$\kpc\ (for
$\Omega_0 = 0.25$) to 266\kms\ at 
0.4\arcsec\ = $1.7\,h_{75}^{-1}$\kpc.  This would be extremely unusual 
behavior for a large elliptical galaxy, not to mention a spectrum
which has been convolved with the seeing and instrumental profile.
We do not confirm their rapid drop in $\sigma$, and do not
agree with their conclusion that the lensing galaxy harbors a massive,
dark object.  We
believe therefore that their results are subject to substantial
systematic error (perhaps because of scattered light), that the
error that they assign to their dispersion measurement is much too
optimistic, and that our measurement is probably more reliable.

Grogin and Narayan (1996) offer two formulae for calculation of $H_0$
from a time delay and a velocity dispersion.  Using the time delay of
$1.142\pm0.004$~year (1-sigma) from Kundi\'c et al. (1997), and a
velocity dispersion of $288\pm9$, the two models give
$72\pm7$\kmsMpc\ (FGS model) and $70\pm7$\kmsMpc\ (SPLS
model).  The velocity dispersion and model give roughly equal
contributions to the error.  It would be quite possible to improve the
accuracy of the velocity dispersion by reobserving the lens at higher
dispersion, perhaps with a 1200~line/mm grating, and it is also
possible that the models need improvement.  For example, 
Bernstein et al. (1997) use HST WFPC2 images of 0957 to 
measure the center of the lensing galaxy, and find that its
location is inconsistent with the GN model.

The main uncertainty in the modeling is due to anisotropies in the
velocity dispersion. Richstone and Tremaine (1984) showed that
the measurement of  velocity dispersions through a central aperture
does not constrain the mass of an elliptical galaxy very well.
From their figures, we estimate the uncertainty in the mass at 40\%
for our aperture of $1\times2.3$ arcsec and
an effective radius of 4.5 arcsec (Bernstein, Tyson, \& Kochanek, 1993).

A recent analysis by Romanowsky \& Kochanek (1998) has confirmed 
this additional uncertainty. They find that the mass cannot
be well measured
if only a central velocity dispersion is measured. They show
that an accurate measurement of the lineprofiles can constrain the
mass much better, but unfortunately, no such measurements are
available for Q0957. Our data give uncertainties on the order
of 0.04 on the Gauss-Hermite terms used for such an analysis
(e.g., van der Marel \& Franx 1993). 
When Romanowsky \& Kochanek impose the constraint that the 
Gauss-Hermite terms do not deviate from mean Gauss-Hermite terms 
measured for local galaxies, they find that the uncertainty in $H_0$ is
on the order of 15 \%.
Given these uncertainties, it is clear that single lenses will not
give a very accurate value for $H_0$, but larger samples may provide
a good mean.

Despite these uncertainties, 
it is heartening to find that we are beginning to converge
on a value for the Hubble constant via very different paths.  Recent
values which depend on the Cepheid calibration include $82\pm8$ (Lauer
et al. 1997, SBF of HST images of BCGs), $81\pm6$ (Tonry et al. 1997,
SBF of nearby galaxies), $73\pm6\pm8$ (Freedman 1997, reporting on the
Cepheid calibration of a variety of secondary distance estimators
including supernovae and Tully-Fisher), 
and $69\pm5$ (Giovanelli et al. 1997, Tully-Fisher relation).  Other
values which are nominally independent of the Cepheid calibration include
$64\pm20$ (Schechter et al. 1997, PG1115 time delay), and $54\pm14$
(Myers et al. 1997, Sunyaev-Zeldovich effect).  This is not the forum
to discuss the relative merits of various methods, but it is clear
measurements of the Hubble constant are finally attaining some
degree of maturity, and that measurements from gravitational
lens time delays can be an important contributor to the quest for $H_0$.

\acknowledgements
We are grateful to P. Fischer for providing us with the exceedingly
deep image of the 0957 field used in Figure 1.  As always, we are
deeply appreciative of the capabilities of the Keck telescopes and the
LRIS spectrograph.

\clearpage

\begin{deluxetable}{lrrrrrr}
\tablecaption{Observing Log.\label{tbl1}}
\tablewidth{0pt}
\tablehead{
\colhead{Objects} & \colhead{UT Date}  &
\colhead{UT}  & \colhead{$\sec\,z$} & 
\colhead{PA} & \colhead{Exposure} & 
\colhead{Slit}
} 
\startdata
AGK2+14783    & 3/30 &  5:39 & 1.01 &  90 &      & 1.0  \nl
AGK2+14783    & 3/30 &  5:46 & 1.01 &  90 &      & 0.7  \nl
0957 GL       & 3/30 &  6:14 & 1.31 & 108 & 1500 & 1.0  \nl
0957 QSO B    & 3/30 &  6:54 & 1.26 & 288 & 1500 & 1.0  \nl
0957 GL       & 3/30 &  7:28 & 1.24 & 108 & 1500 & 1.0  \nl
0957 QSO B    & 3/30 &  7:59 & 1.24 & 288 & 1500 & 1.0  \nl
0957 QSO A    & 3/30 &  8:26 & 1.24 & 288 &  500 & 1.0  \nl
MS1358 G1+... & 3/30 & 11:12 & 1.37 &  24 & 1500 & 1.0  \nl
MS1358 G1+... & 3/30 & 11:39 & 1.36 &  24 & 1500 & 1.0  \nl
HD132737      & 3/30 & 15:23 &      &  90 &      & 0.7  \nl
HD132737      & 3/30 & 15:27 &      &  90 &      & 1.0  \nl
AGK2+14783    & 3/31 &  5:14 & 1.00 &  90 &      & 1.0  \nl
AGK2+14783    & 3/31 &  5:26 & 1.01 &  90 &      & 0.7  \nl
0957 GL       & 3/31 &  6:55 & 1.25 & 108 & 1500 & 1.0  \nl
0957 GL       & 3/31 &  7:25 & 1.24 & 108 & 1500 & 1.0  \nl
\enddata
\tablecomments{PA is east from north, exposures are in seconds,
slit widths are in arcseconds.}
\end{deluxetable}

\begin{deluxetable}{lrrrrrrrrrrl}
\tablecaption{Redshifts and Dispersions.\label{tbl2}}
\tablewidth{0pt}
\tablehead{
\colhead{ID} & \colhead{Y\#} & \colhead{G\#} & \colhead{A\#} & 
\colhead{$-\alpha$(\arcsec)} & \colhead{$\delta$(\arcsec)} & 
\colhead{$z$} & \colhead{$\pm$} & \colhead{$\sigma$} & \colhead{$\pm$}
& \colhead{$r$} & \colhead{emission}
} 
\startdata
G1 & 88&   &  & $ -0.3$ & $  1.1$ & 0.3562 & 0.0001 & 294 &  10 &  12.6 &  \nl
t1 & 40&R33&  & $ 60.9$ & $ 20.4$ & 0.5044 & 0.0005 & 193 &  70 &   4.7 &  \nl
t2 &100&   &21& $-11.2$ & $ -2.6$ & 0.3550 & 0.0001 & 105 &  66 &  14.6 &  \nl
t3 &115&   &  & $-22.5$ & $ -6.3$ & 0.3583 & 0.0001 &  72 & 100 &  14.6 &  \nl
t4 &164&   &  & $-79.0$ & $-24.7$ & 0.9120 & 0.0003 &  &  &  & [OII] \nl
t5 &116&R19&  & $-22.8$ & $ -8.3$ & 0.3532 & 0.0001 & 154 &  63 &   8.3 &  \nl
t6 & 97&R13&20& $ -7.5$ & $  3.1$ & 0.3547 & 0.0002 &  68 & 143 &   6.9 &  \nl
t7 &   &   &  & $ 21.6$ & $  8.9$ & 0.4476 & 0.0004 &  &  &  & [OII],$H_\beta$,[OIII] \nl
t8 &   &   &  & $ 27.4$ & $ 13.9$ & 0.9070\rlap{?} & 0.0005 &  &  &  & [OII]\rlap{?} \nl
\enddata
\tablecomments{Columns: 
Galaxy name, Identification numbers from Young et al. (1981), Garrett et
al. (1992), and Angonin-Willaime et al. (1994), Position with respect
to QSO component B (arcsec), redshift and error, velocity dispersion
(\kms) and error, cross-correlation $r$ value, and emission lines
observed.}
\end{deluxetable}

\endinsfig

\clearpage

\centerline{\bf FIGURE CAPTIONS}
\bigskip

\figcaption[f1.eps]{
Illustration of the slit positions and identifications of the galaxies
observed here.  North is up and East is left; G2--G5 are the
designations of Young et al. (1980); and \#6 is galaxy 20 and \#2 is
galaxy 21 from Angonin-Willaime et al. (1994).
\label{fig1}}

\figcaption[f2.eps]{ The spectra are shown for the four galaxies at
redshift 0.355, t6, t5, t3, and t2 (offset by 400, 250, 150, and 0 for
clarity), and the four galaxies at higher redshift, t1, t4, t7, and t8
(offset by 800, 400, 150, and 0).
These spectra have been Gaussian smoothed with a \FWHM\ of
6\AA.  The top axis shows rest wavelength at a redshift of 0.355.
\label{fig2}}

\figcaption[f3.eps]{The combined spectrum of the lensing galaxy G1 and
the leakage of light from QSO component B are shown, as well as the
scaled spectrum of QSO component A and the difference which is the
``pure'' spectrum of G1 which is subjected to redshift and dispersion
analysis.  These spectra have been Gaussian smoothed with a \FWHM\ of
6\AA.  The top axis shows rest wavelength at a redshift of 0.355.
\label{fig3}}

\figcaption[f4.eps]{The normalized spectrum, the fit derived from the 
Fourier fit (offset by $-0.15$), and the difference (offset by +0.5)
are plotted as a function of observed wavelength.  The big feature at
5840\AA\ is the G band.  The quality of the fit is clearly very good.
\label{fig4}}

\figcaption[f5.eps]{The velocity dispersion is plotted as a function of
position along the slit (open symbols are results from use of stellar
templates; star symbols show results from line fitting).  The data
points from FSMD are also shown as small, filled symbols.
\label{fig5}}


\clearpage
\begin{thebibliography}{}


\bibitem[Angonin 1994]{ang94}
 Angonin-Willaime, M.-C., Soucail, G., \& Vanderriest, C. 1994, 
  \aap, 291, 411.

\bibitem[Bernstein 1994]{btk94}
   Bernstein, G., Tyson, J.A., \& Kochanek, C.S. 1993,
   \aj, 105, 815.

\bibitem[Bernstein 1997]{ber97}
   Bernstein, G., Fischer, P., Tyson, J.A., \& Rhee, G. 1997
   \apjl, 000, 000.

\bibitem[Falco 1996]{fal96}
   Falco, E.E., Shapiro, I.I., Moustakas, L.A., \& Davis, M. 1996,
   \apj, 000, 00.

\bibitem[Fischer 97]{fis97}
  Fischer, P., Bernstein, B., Rhee, G., \& Tyson, J.A. 1997, \aj, 113,
  421.

\bibitem[Freedman 96]{fre96}
  Freedman, W.L. 1996 18th Texas Symposium, December 1996, 
  to be published by World Scientific, 
  eds. A. Olinto, J. Friemann and D. Schramm.

\bibitem[Grogin 96]{gro96}
  Grogin, N.A. \& Narayan, R 1996 \apj, 464, 92, and 473, 570.

\bibitem[Garrett 92]{gar92}
  Garrett, M.A., Walsh, D., \& Carswell, R.F. 1992, \mnras, 254, 27p.

\bibitem[Haarsma 1997]{haa97}
 Haarsma, D.B.; Hewitt, J.N., Lehar, J., \& Burke, B.F. 1997, \apj, 497, 102.

\bibitem[Franx 1989]{fra89}
 Franx, M., Illingworth, G.D., \& Heckman, T.M. 1989 \apj, 344, 613.

\bibitem[Franx 1993]{fra93}
 Franx, M. 1993 \apjl, 407, 5.

\bibitem[Giovanelli 1997]{gio97}
 Giovanelli, et al. 1997 \apjl, 477, 1.

\bibitem[Kelson 1997]{kel97}
 Kelson, D.D., Illingworth, G.D. Franx, M., \& van Dokkum, P.G.
 1997, \apj, in preparation.
 
\bibitem[Kundic 1997]{kun97}
 Kundi\'c, T., Turner, E.L., Colley, W.N., Gott, J.R., Rhoads, J.E., 
Wang, Y., Bergeron, L.E., Gloria, K.A., Long, D.C., Malhotra, S., \&
Wambsganss, J.
1997, \apj, 482, 75.

\bibitem[Lauer 1997]{lau97}
  Lauer, T.R., Tonry, J.L., Postman, M., Ajhar, E.A., \& Holtzman,
  J.A. 1997, \apj, 000, 000.

\bibitem[Marel 1994]{mar94}
  van der Marel, R., \& Franx, M. 1993, \apj, 407, 525.

\bibitem[Myers 1997]{mye97}
  Myers, S.T., Baker, J.E., Readhead, A.C.S., Leitch, E.M., \& Herbig
  T. 1997, \apj, 000, 000.

\bibitem[Oke 1995]{oke95}
 Oke, J.B., et al. 1995, \pasp, 107, 375.

\bibitem[Osterbrock 1996]{ost96}
 Osterbrock, D.E., Fulbright, J.P., Martel, A.R., Keane, M.J., \&
 Trager, S.C. 1996, \pasp, 108, 277

\bibitem[Press 1992]{pre92}
 Press, W.H., Rybicki, G.B., \& Hewitt, J.N. 1992, \apj, 385, 404.

\bibitem[Refsdal 1964]{ref64}
  Refsdal, S. 1964 \mnras, 128, 307.

\bibitem[Romanowsky 1998]{rom98}
 Romanowsky, A. J., \& Kochanek, C. S., 1998, preprint, astroph/9805080.

\bibitem[Richstone 1984]{rich84}
 Richstone, D. O., \& Tremaine, S. D. 1984, \apj, 286, 27.

\bibitem[Sargent 1977]{sar77}
 Sargent, W.L.W., Schechter, P.L., Boksenberg, A., \& Shortridge,
 K. 1977, \apj, 212, 326.

\bibitem[Schechter 1997]{pls97}
 Schechter, P.L. et al. 1997, \apjl, 475, L85.

\bibitem[Schild 86]{sch86}
  Schild, R.E., \& Cholfin, B. 1986, \apj, 300, 209.

\bibitem[Tonry 1979]{ton79}
  Tonry, J.L. \& Davis, M. 1979, \aj, 84, 1511.

\bibitem[Tonry 1984]{ton84}
  Tonry, J.L. 1984, \apj, 279, 13.

\bibitem[Tonry 1997]{ton97}
  Tonry, J.L., Blakeslee, J.P., Ajhar E.A., \& Dressler, A. 1997, 
  \apj, 475, 399.

\bibitem[Tonry 1998]{ton98}
  Tonry, J.L. 1998, \aj, 115, 1.

\bibitem[van Dokkum 1996]{van96}
  van Dokkum, P.G., \& Franx, M. 1996, \mnras, 281, 985.

\bibitem[Young 1980]{you80}
  Young, P., Gunn, J.E., Kristian, J., Oke, J.B., \& Westphal,
  J.A. 1980, \apj, 241, 507.

\bibitem[Young 1981]{you81}
  Young, P., Gunn, J.E., Kristian, J., Oke, J.B., \& Westphal,
  J.A. 1981, \apj, 244, 736.




\end{thebibliography}
\end{document}